# Thermodynamically driven tilt grain boundaries of monolayer crystals using catalytic liquid alloys

*Min-Yeong Choi[1,2,†], Chang-Won Choi[1,3,†], Dong-Yeong Kim[1,2], Moon-Ho Jo[1,3], Yong-Sung Kim[4,5,\*], Si-Young Choi[1,3,6,\*], and Cheol-Joo Kim[1,2,\*]*

[1]Center for Van der Waals Quantum Solids, Institute for Basic Science (IBS), Pohang 37673, Republic of Korea

[2]Department of Chemical Engineering, Pohang University of Science and Technology (POSTECH), Pohang 37673, Republic of Korea

[3]Department of Materials Science & Engineering, POSTECH, Pohang 37673, Republic of Korea

[4]Korea Research Institute of Standards and Science, Daejeon 34113, Republic of Korea

[5]Department of Nano Science, University of Science and Technology, Daejeon 34113, Republic of Korea

[6]Department of Semiconductor Engineering, POSTECH, Pohang 37673, Republic of Korea

[\*]Correspondence to: yongsung.kim@kriss.re.kr, youngchoi@postech.ac.kr, kimcj@postech.ac.kr

[†]These authors contributed equally to this work.

**ABSTRACT**

We report a method to precisely control the atomic defects at grain boundaries (GBs) of monolayer $MoS_2$ by vapor–liquid–solid (VLS) growth using sodium molybdate liquid alloys, which serve as growth catalysts to guide the formations of the thermodynamically most stable GB structure. The Mo-rich chemical environment of the alloys results in Mo-polar 5|7 defects with a yield exceeding 95%. The photoluminescence (PL) intensity of VLS-grown polycrystalline $MoS_2$ films markedly exceeds that of the films exhibiting abundant S 5|7 defects, which are kinetically driven by vapor–solid–solid growths. Density functional theory calculations indicate that the enhanced PL intensity is due to the suppression of non-radiative recombination of charged excitons with donor-type defects of adsorbed Na elements on S 5|7 defects. Catalytic liquid alloys can aid in determining a type of atomic defect even in various polycrystalline 2D films, which accordingly provides a technical clue to engineer their properties.

Structural defects in a polycrystalline two-dimensional (2D) structure can often reform the material properties beyond those of single crystals[1]. In particular, grain boundaries (GBs) contain various atomic defects that exhibit novel electrical[2,3], magnetic[4,5], mechanical[6,7], and chemical[8–10] properties. The emergent properties of GBs are extremely sensitive to atomic arrangements with short-range ordering, which are thermodynamically metastable and thus can have variable configurations. To obtain desirable properties associated with specific structural defects, GB structures must be elaborated upon at the atomic level. Yet, polycrystalline 2D structures have exhibited uncontrolled GBs in which various defect structures coexist[11,12].

The structures of tilt GBs are primarily characterized by two structural parameters: the tilt angle, $\theta_t$, and the inclination angle, $\Phi$, of defects within the lattice (Figure 1a). In transition metal dichalcogenides ($MX_2$; M = Transition metal, X = Chalcogen) such as $MoS_2$ with three-fold rotational symmetry, the magnitude of $\theta_t$ varies from 0° to 60°, and the defect density increases as $\theta_t$ increases (Figure S1 for atomic configurations as a function of $\theta_t$). While it is energetically favorable to form pentagon–heptagon (5|7) defects with the smallest Burgers vector, $\vec{b}$, considering the strain energy of GBs[13], the 5|7 defects have a structural degree of freedom to exhibit two types of homoelemental bonding (HBs), either M–M or X–X, depending on $\Phi$, which is associated with the crystallographic orientation of defects[11,14]. Here, we define $\Phi$ by the relative rotational angle between the defect direction vector, $\vec{x}$ (dotted blue arrow) and the symmetric GB vector, $\vec{v}$ (solid black arrow), where $\vec{x}$ is parallel to the vertical mirror plane of the defects, pointing to the seven ring from the five ring, and $\vec{v}$ is aligned to the symmetric armchair (AC) direction between the two tilted zig–zag (ZZ) axes (dotted lines) of the merged grains. The defects with $\Phi$ = 0° form 5|7 membered rings with Mo–Mo HB (Mo 5|7), whose $\vec{b}$ is equal to $\vec{a}_1$. On the other

hand, the defects with $\Phi = 60°$ result in S 5|7 structures with S–S HB that have $\vec{b}$ of $\vec{a}_2$. In theory, the formation of specific defects can be thermodynamically guided by modulating the chemical potential of the environment[15]. While the Mo 5|7 defect is more stable than the S 5|7 defect in a Mo-rich environment with low S chemical potential, $\mu_s$, the S 5|7 defect becomes more stable than Mo the 5|7 defect in a Mo-poor environment with high $\mu_s$ (Figure 1b).

Although individual control of M- and X-containing precursors during vapor-phase growth[16,17] would be the simplest way of changing $\mu_s$ to determine the thermodynamically most stable defect structures, two main factors prevent this approach from realizing the deterministic formation of specific atomic defects. First, owing to the spatiotemporal fluctuations of introduced gaseous precursors, developing uniform environments across a large area is challenging. If a large amount of either M or X elements is supplied to induce an extreme chemical environment for selective defect formations that resist the fluctuation of vapor flows, the excess elements can produce undesired byproducts, such as solid M or X and other compounds[18,19]. Second, the final GB structures are often set kinetically. When the continuous growth of adjacent grains tilted by a specific $\theta_t$ "zips-up" the gap between joint grains, the geometric boundary condition can limit possible $\Phi$, resulting in thermodynamically unfavorable defect structures under the given chemical condition[20,21] (Figure S2 for the schematics of the process). To laterally connect tilted grains with defects of the same $\Phi$, GB migration by reordering atomic structures is required to overcome a significant energy barrier[22]. Notably, in previous studies, even growth with a controlled chemical environment resulted in polycrystalline $MoS_2$, where two types of different GBs with orientations tilted by roughly 60° coexisted at the joint of merged grains[20,23], suggesting the kinetically driven formation of GBs[21].

We propose a synthetic strategy to utilize catalytic alloys of sodium molybdate, Na–Mo–O, in metal–organic chemical vapor deposition (MOCVD)[16]. The eutectic composition[24] of a high Mo concentration of about 20 mol% can provide a stable, low $\mu_s$ environment with a large amount of chemically active Mo elements in a liquid phase[25], via which vapor–liquid–solid (VLS) growth occurs. In addition, the Na element in the liquid alloys functions as a catalyst to lower the energy barrier for the transformation of the kinetically driven other defect structures to Mo 5|7 defects[26]. This approach is possible due to the 2D structure of the material, where all the GBs are exposed to the surface to interact with the catalytic alloys.

Monolayer $MoS_2$ was grown with and without liquid alloys in a single batch to test their roles in GB defect formation (Figure 1c, detailed methods in Supporting Information)[27]. The GBs of the two polycrystalline $MoS_2$ films grown by vapor-solid-solid (VSS) and VLS growth, respectively, were directly visualized using a scanning transmission electron microscope with a high-angle annular dark field detector (HAADF-STEM). Tilted grains of $\theta_t$ formed by VSS and VLS growths are shown in the inverse fast Fourier transform (FFT) images of the STEM image data by selecting reciprocal points, where the ZZ planes of each grain are color-coded differently (Figure 2a, 2b). Notably, the GB was nearly aligned to the symmetric axis of $\vec{v}$ in the VLS growth, whereas it was rotated roughly by 30° with respect to $\vec{v}$ in the VSS growth. In the HAADF-STEM data at the atomic scale (Figure 2c, 2d), defects with two different $\Phi$, close to either 0° (highlighted with purple ⊥) or 60° (red ⊥) coexisted in the VSS growth. However, although the entire GB was still asymmetric, forming a nonstraight line, $\vec{x}$ of all defects was closely parallel to $\vec{v}$ with $\Phi \sim 0°$ in the VLS growth (Figures S3 and S4 for STEM data over different GB areas). The statistics for $\Phi$ values and atomic structures of defects is shown for both VSS and VLS growths from 150 defects in total (Figure 2e). VLS growth resulted in Mo 5|7 defects with a yield > 95% along the GBs, with the defects exclusively pointing to $\Phi \sim 0°$. However, VSS growth resulted in varying defects.

Changes of the defect structures by incorporating interstitial or substitutional elements (Figure 2f, 2g) can happen easily, while maintaining the $\Phi$ due to the low activation energies[15]. Because the local chemical environment can vary over time during VSS growth, the atomic defects can translate into the new structure, resulting in various derivatives for both Mo 5|7 and S 5|7 defects[14]. The formation of Mo 5|7 and S 5|7 defects were expected to occur in completely different regimes of $\mu_s$ (Figure 1b), implying that their coexistence indicates a large temporal variation in local chemistry during vapor-phase growth. However, during VLS growth, Mo 5|7 defects formed at the low $\mu_s$ were preserved by the nearby liquid alloys after stitching of grains. Because the alloys comprise more Mo and S elements than in gaseous environments, the chemical potential within the alloys can be mostly maintained despite temporal vapor pressure fluctuations[25].

Meanwhile, simply modulating the growth environment without catalytic reaction would kinetically limit GB structures, resulting in both 0˚ and 60˚ for the $\Phi$ values[11]. If the energy barrier for the growth kinetics of kink nucleation and propagation is high[21], the rough GBs, even at the nanometer scale, can be formed with multiple kinks and various defects of different $\Phi$ values. However, in our VLS growth, although merging two adjacent edges of tilted grains with Mo 5|7 defects were geometrically restricted at local areas, $MoS_2$ near the GBs could be continuously dissolved in the liquid alloys and then resolidified, prompting the reordering of the crystalline structures with Na catalysts[26] to form the thermodynamically most stable Mo 5|7 defect with $\Phi \sim$ 0˚. Indeed, we observed smooth GBs without noticeable kinks in VLS growth, which are

thermodynamically expected forms with minimized total lengths of GBs between two tilted grains to lower the structure's total energy[28].

Now, we discuss how the GB structures change as a function of $\theta_t$. We deduce the density of defects, $d_{defects}$, along GBs of different $\theta_t$ by the inverse value of the averaged distances between defects with the same $\Phi$. In both samples grown by VLS and VSS modes, $d_{defects}$ increased monotonically as $\theta_t$ increased in a quantitatively similar manner (Figure S5). The change in $\theta_t$ was closely matched with the model for the maximum relaxation of strains at the tilt GB by formations of defects (Supporting Information for the tilt-boundary model)[29]. At GBs with $\theta_t \sim 34°$, $d_{defects}$ became large enough to form closely packed defects with $\Phi \sim 0°$ in series (Figure S3). Because accommodating a larger $d_{defects}$ with a single type of defects at a tilt GB with $\theta_t > 34°$ is difficult, theories predict the formation of alternating Mo 5|7 and S 5|7 defects, pointing to different orientations for relaxing the local strains at the GBs[4,6]. Therefore, when stitching grains with a single type of defect structure in polycrystalline films, high-$\theta_t$ GB formation must be avoided. The $MoS_2$ grains are grown on amorphous surfaces, therefore their crystallographic orientations should be randomly defined, meaning that $\theta_t$ values between merging grains can have any values up to 60°, including high $\theta_t$. Nevertheless, we did not observe high-$\theta_t$ GBs with $\theta_t > 34°$, indicating a self-exclusion of the formations of high-$\theta_t$ GBs.

We investigated the regions near two grains with a high misorientation angle of 46°. An atomic STEM image is overlayed with its false-color inverse FFT image, which presents the ZZ planes of two tilted grains, corresponding to reciprocal points from each grain (Figure 3a, 3b). We found that if a certain reciprocal point was selected for the inverse FFT from the *A* grain, the resultant image showed a reduced intensity along the linear region within the grain, separating the grain into *A* and *A*\* grains. The reduced intensity indicated a missing ZZ plane corresponding to the selected reciprocal point. In the atomic image (Figure 3c) across the boundary regions (□ region in Figure 3b), we found that the inverse *A* and *A*\* grains are translated slightly along the ZZ direction to form a mirror-twin boundary (MTB) with a one-dimensional chain of edge-shared 4|4 defects (4|4 E), and a low tilt angle, $\theta_{t\text{-}A^*B}$, was formed between the *A*\* grain and *B* grain. The tilt angle between the *A* grain and *B* grain $\theta_{t\text{-}AB}$ of 46° was equivalent to the sum of $\theta_{t\text{-}AA^*}$ and $\theta_{t\text{-}A^*B}$. The $\theta_{t\text{-}AA^*}$ for MTB was 60° or equivalent to 180°, considering the three-fold rotational symmetry, and $\theta_{t\text{-}A^*B}$ was −14°; the negative sign means that the tilt angle polarity for $\theta_{t\text{-}A^*B}$ is opposite to that of $\theta_{t\text{-}AA^*}$. Meanwhile, the defects at the GB between the *A*\* grain and *B* grain form Mo 5|7 defects. GB-mediated deformation twinning in other regions is also presented in Figure S6.

To understand the deformation–twinning process, we present a shear strain, $\gamma_{shear}$ map in Figure 3d (○ region in Figure 3b). The $\gamma_{shear}$ is deduced by *tan* (*k*), where *k* indicates the local distortion angle of the ZZ axis toward the clockwise direction with respect to the perpendicular direction of the AC axis from the image of Mo atoms (Figure 3e). High local strains of $|\gamma_{shear}|$ ~ 0.2 with opposite polarities present near the intersection between MTB and low-$\theta_t$ GB, where two grains with high $\theta_t$ are directly connected. In contrast, the strains show small values < 0.03 in other regions, including the areas near MTB and low-$\theta_t$ GB, indicating that the deformation twinning can effectively relax the strain at high-$\theta_t$ GBs (Figure 3f for the atomic images of the high-strained and low-strained regions, which are highlighted in Figure 3d).

Our hypothesis for the deformation–twinning process that formed the structure in Figure 3d is illustrated in Figure 3g. When titled *A* and *B* grains with a high $\theta_{t\text{-}AB}$ were merged, closely packed single-type defects were formed to stitch the two grains (Figure 3g, step *i*). Due to the low $\mu_s$ environment, given by the liquid alloys, Mo 5|7 defects inclined to $\vec{v}_{AB}$ with $\Phi \sim 0°$ were selectively formed. Because the $d_{defect}$ for the single type of Mo 5|7 defect was limited, high residual strains were accumulated near the GB with Mo HBs, which were the weakest bond in the lattice (Figure S7 and discussion for the structural analyses of defects in Supporting Information). The $\gamma_{shear}$ above the critical value cause structural instability. As discussed earlier, forming additional S 5|7 defects provides a way to release the strain; however, S 5|7 defects are thermodynamically unfavorable at low $\mu_s$. Another possible structural transformation to release the strain is the lateral gliding of the S planes in one of the grains (step *ii*). If the gliding of S planes is pinned to a certain point (step *iii*), an MTB with 4|4 E defects (highlighted by gray colors) is formed[30]. The deformation–twinning process relax the total interfacial and strain energy. MTB lacks energetically unfavorable HB in terms of chemical energy. The emerging low-tilted boundary (LTB) has HB, but its $d_{defects}$ is significantly lower than that of the high tilt GB. Therefore, when an MTB nucleates by a critical strain field at an unstable tilted GB, it can spontaneously grow by inducing cooperative displacement of S atoms, resulting in twinning with plastic deformation. After the process, the strain field only remains at the intersection between coherent MTB and incoherent, curved LTB[31]. Notably, the absorption and emission of defects during the deformation–twinning process in VLS growth selectively left Mo 5|7 defects as a stable GB structure.

To identify the roles of different defects in determining the physical properties of polycrystalline materials, we further compared the photoluminescence (PL) near the GBs formed by VLS and VSS growths (Figure 4). For the direct comparison, we devised a growth method by which VLS- and VSS-GBs with a similar $\theta_t$ can be simultaneously generated in a single sample. Here, the growth mode was switched from the VLS mode to the VSS mode on SL substrates by lowering the temperature during the growth below the melting point of Na-Mo-O alloys, $T_m \sim 500\ °C$ (Figure 4a). In STEM images, we confirmed that the edges of the first-grown region (Figure 4a, blue dotted line and box) and the second-grown region (red dotted line and box) of the single crystal showed ZZ–Mo and ZZ–S edges, respectively, which were expected for low and high $\mu_s$ for each growth mode[32] (see Figure S8 and Supporting Information for the experimental details). By the two-step growth, titled grains could be stitched either by VLS or VSS growths, generating VLS- (Figure 4a, highlighted by solid blue line) and VSS-GBs (solid red line), respectively.

The VLS- and VSS-grown regions were clearly distinguished in the PL intensity mapping images of the as-grown samples (Figure 4b), where the PL intensity near the triangular boundaries between the VLS- and VSS-grown regions was significantly quenched by the solidified alloys along the boundaries. We observed that the PL spectra from the grain interior (GI) showed similar intensities in both VLS- and VSS-grown regions, suggesting that the growth mode did not affect the GI's atomic structures. Considering the relative angle between the lines with quenched PL in the merged grains, we deduced $\theta_t$. Representative PL mapping data (Figure 4c) near the tilt, VLS- and VSS-GBs of $\theta_t \sim 33°$, showed different behaviors. The PL intensity at the VLS-GB (the region indicated by ◊ symbol in Figure 4c) was similar to that from the nearby GI. In contrast, at VSS-grown GBs (the region indicated by ▽ symbol in Figure 4c), the PL peak was red-shifted by ~10 meV with significant suppression of intensities by as much as ~ 55 % at the incident power density of ~$5 \times 10^3$ W/cm$^2$ (Figure 4d). In the measurements on different samples, VLS-GBs always showed similar or even higher PL intensity than GIs, whereas VSS-GBs often showed lower intensities than GIs. We also measured the global PL over $30 \times 30$ μm$^2$ are in the continuous films with polycrystalline structures, which we synthesized by VLS and VSS growth modes, separately. The PL spectra showed significantly stronger intensities in the VLS-grown films than in the VSS-grown films (Figure S9). By spectral deconvolution, the negative trion peak area $I_{tr}$ and exciton peak area $I_{ex}$ were deduced, and the ratio of the spectral weights $w$ was calculated as $I_{tr}/I_{ex} = 0.28$ for VLS-grown films and 0.46 for VSS-grown films (Figure S9). The result suggests the spontaneous formation of negative trions with the existence of electron doping concentration, $n$, which have a lower efficiency for radiative recombination than do excitons[33,34].

We conducted density functional theory (DFT) calculations to understand how the main four defects observed in our samples (*i.e.,* Mo 5|7, Mo 6|8, S 5|7, and S 4|6 defects) changed $n$. The calculations show that all the calculated defects form deep-level defect states near mid-gap, which cannot contribute to an effective electron doping at room temperature (Figure S10). Then, we speculated that the defects host other donor-like impurities. According to our calculations, Na atoms interact more strongly to the Mo 5|7, Mo 6|8, S 5|7, and S 4|6 defects than the pristine lattice, presumably due to their strong tendency to capture electrons from the alkali metal. All the defect structures can spontaneously host two Na atoms, as depicted in the schematics of Figure 4e. We calculated $E_F$ for GB loop structures with atomic defects and different numbers of Na elements (Figure S11) and compare the level with $E_F$ for the pristine lattice (Figure 4f) to estimate the efficiency of the adsorbed Na atoms for doping of free electrons to the $MoS_2$ lattice. With Mo 5|7, Mo 6|8, and S 4|6 defects, $E_F$ are located at deep level below the conduction band edge of the pristine lattice by ~1 eV with two surrounding Na atoms. However, for S 5|7 defects, the position of $E_F$ moves within the conduction band (Figure 4g). The electron trap efficiency of the atomic defects is related to the density of trap states of the defects. S 5|7 defects have fewer mid-gap states

to trap free electrons than other Mo 5|7, Mo 6|8, and S 4|6 defects. Therefore, S 5|7 defects cannot fully accommodate the electrons donated by Na atoms, keeping the free electrons from Na atoms in the $MoS_2$ lattice, while other defects effectively trap all the electrons. We note that the defects could host other impurities to change the electron doping level as well[35], but we suspect that the electron trap efficiency would be still higher near Mo 5|7, Mo 6|8, and S 4|6 defects for the same reason mentioned above. We have further conducted quantitative comparisons between $n$ values, deduced from the theory and the experimentally measure PL, and they showed reasonable agreements (Supporting Information).

In conclusion, we demonstrated that using catalytic liquid alloys can provide a powerful approach with unprecedented precision for defect control at GBs by thermodynamics. We showed that the electrical doping level of the 2D films significantly depends on the type of defects at GBs due to their different interactions with foreign elements. Our technique can be broadly applied to engineering various electrical, chemical, and mechanical properties of 2D materials, realizing novel structures such as one-dimensional quantum wire[3], single atom catalysts[36], and extremely flexible membrane[37]. Finally, if appropriate alloys can be developed, this approach is versatile enough to produce various atomic defects in a library of 2D materials.

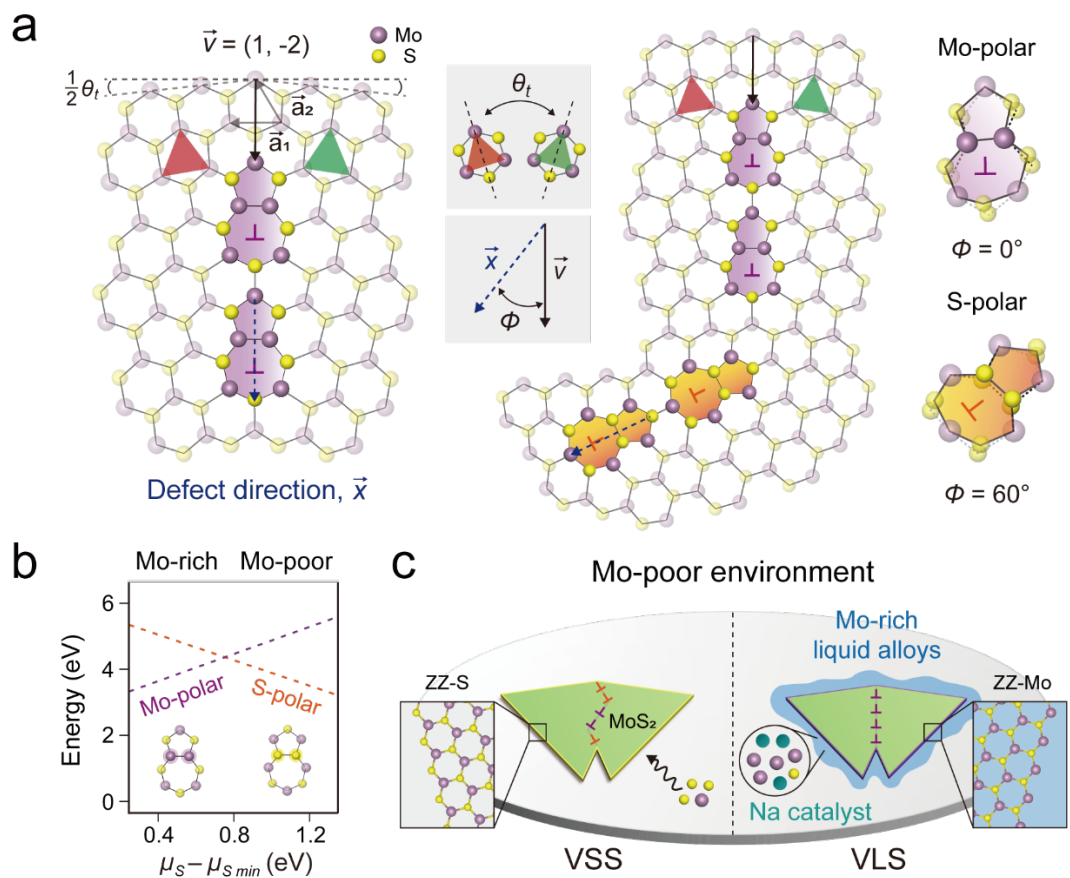

**Figure 1**. **Growth-mode-dependent GBs.** (a) Schematics of the formation of tilt boundaries. Stitching of grains tilted by $\theta_t$ can form Mo 5|7 defects with $\Phi = 0°$ and S 5|7 defects with $\Phi = 60°$. (b) Density functional theory (DFT) calculations for the formation energy of GB defects with Mo–Mo or S–S HB as a function of $\mu_S - \mu_{S\,min}$ of the growth environment (see Supporting Information for the definition of $\mu_{S\,min}$). (c) Illustration of vapor–solid–solid (VSS) and vapor–liquid–solid

(VLS) growth modes with different chemical environments, which result in a kinetically driven mixture of Mo 5|7 and S 5|7 defects and a thermodynamically driven array of homogenous Mo 5|7 defects, respectively.

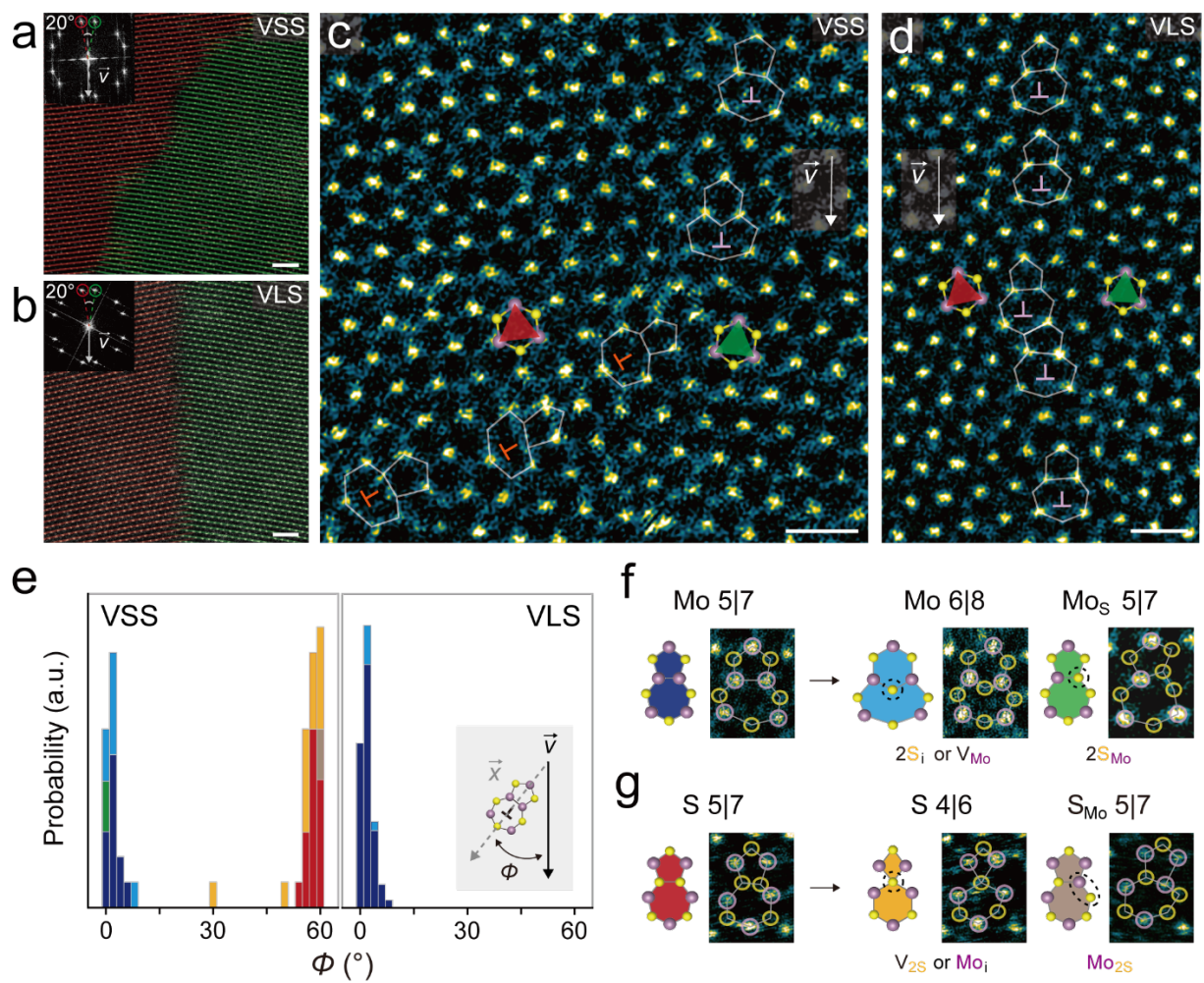

**Figure 2. Comparison of GB structures between VSS and VLS growth modes.** (a, b) False-color fast Fourier transform (FFT)-filtered STEM images of GB structures with a $\theta_t$ of 20° formed by (a) VSS mode and (b) VLS mode. (Inset) Diffraction pattern images show two spots (red and green circles) with a 20° rotation and $\vec{v}$ aligned to the symmetric direction of the two merged crystalline grains. Scale bar: 1 nm. (c, d) Magnified atomic images of each GB structure in 2a and 2b. Scale bar: 0.5 nm. (e) Statistics of $\Phi$ and the types of atomic structures for defects at tilt GBs formed by each growth mode. Here, the $\Phi$ values are defined between 0° and 60°, considering the three-fold rotational symmetry ($\Phi = \Phi + 120°$) and the in-plane mirror symmetry ($\Phi = -\Phi$) of the MoS$_2$ crystal. (f, g) Atomic reconstruction processes of (f) Mo 5|7 and (g) S 5|7 defects to their derivatives. Mo 5|7 derivatives occur at relatively high $\mu_s$ resulting in Mo 6|8 by S interstitials or Mo vacancies and Mo$_S$ 5|7 by 2S substitution for the Mo sites of HB. S 5|7 derivatives include S 4|6 by double S vacancies or Mo interstitials, formed in relatively low $\mu_s$ and S$_{Mo}$ 5|7 with a Mo switched positions with S atoms.

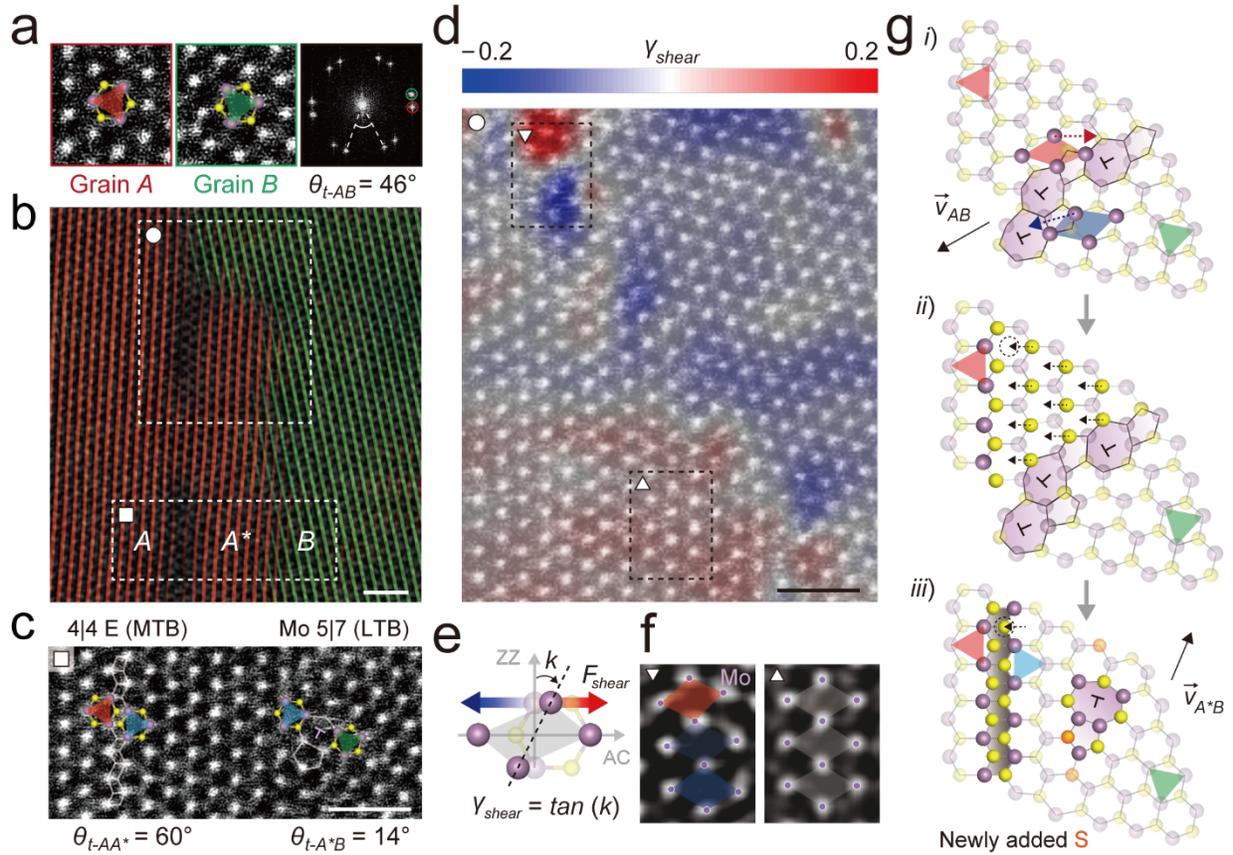

**Figure 3. Deformation–twinning process at high $\theta_t$ stitching.** (a) STEM images and corresponding diffraction pattern of closely located *A* and *B* grains, in which the in-plane crystallographic orientations are rotated by 46°. (b) Combined atomic image and false-color FFT-filtered image. The FFT image was obtained by selecting two spots (red and green circle) of diffraction pattern in Figure 3a. Scale bar: 1 nm. (c) Zoomed-in image over the dotted area in Figure 3b (□), showing three (*A, A\*, B*) regions with different crystallographic orientations. $\theta_{t\text{-}AA^*}$: tilt angle between *A* and *A\** regions; $\theta_{t\text{-}A^*B}$: tilt angle between *A\** and *B* regions. Scale bar: 1 nm. (d) The $\gamma_{shear}$ map, combined with the atomic image at the dotted area in Figure 3b (○). Scale bar: 1nm. (e) A schematics to describe shear strain, $\gamma_{shear}$ from the distorted MoS$_2$ unit cell. Shear force applied toward the AC direction generates the distortion angle, *k*. The $\gamma_{shear}$ is defined as the tangent

value of *k*. (f) Zoom-in STEM images of the high-strained (▽) and low-strained regions (△) from Figure 3d. The local structural distortions are presented by different colors, as described in Figure 3e. (g) Schematics for the deformation–twinning process to generate the GB structure in Figure 3a–f.

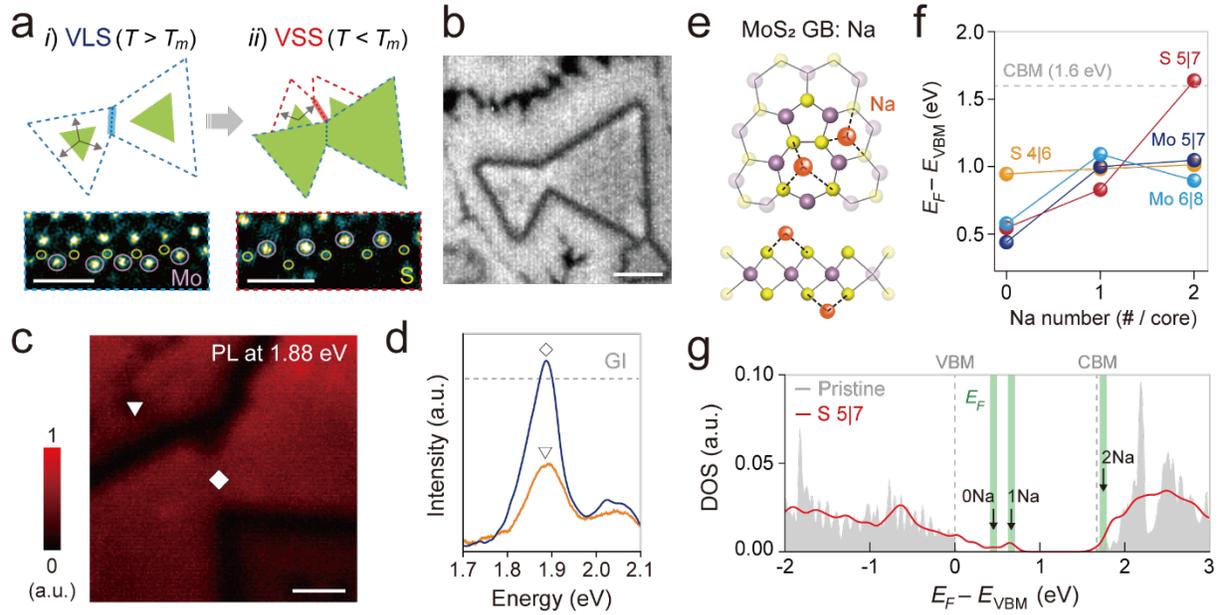

**Figure 4. PL properties of VLS- and VSS-grown GBs.** (a) Consecutive VLS and VSS growth for direct comparison of each GB with a similar $\theta_t$. Inset: STEM atomic images for edges grown by each growth mode. Scale bar: 0.5 nm. (b) PL intensity mapping image for resultant sample grown by the method shown in 4a. Scale bar: 5 μm. (c) PL intensity mapping image for VLS- and VSS-GBs with an excitation energy of 1.88 eV and incident power density of ~ 5 × 10$^3$ W/cm$^2$. Scale bar: 1 μm. (d) PL spectra corresponding to 4c. The dotted lines indicate the intensity measured at the grain interiors. (e) Schematics of top- and side-view atomic structures of Na adsorbed at a defect of MoS$_2$ GBs. (f) Change of the Fermi level ($E_F$) for the structure in Figure 4e with respect to the valence band maxima (VBM) of the pristine lattice as a function of the number of Na atoms adsorbed on different GB defects. (g) Electronic structure of S 5|7 defects embedded in monolayer MoS$_2$. The gray plot indicates the density of state (DOS) for pristine MoS$_2$ without defects, and the red plot indicates DOS for the defects. The gray dotted lines show VBM and the conduction band minima (CBM) for the pristine lattice. The highlighted green lines

indicate the levels to be occupied by the electrons from Na atoms. As the adsorbed Na atoms increase, the occupation levels also increase.

## Supporting Information

**Growth of monolayer MoS$_2$**

Monolayer MoS$_2$ was grown on glass substrates in a home-built metal–organic chemical vapor deposition setup that had a hot-wall furnace. The Mo precursor was Mo(CO)$_6$, and the S precursor was (C$_2$H$_5$)$_2$S. The existence of liquid alloys during growth can be determined by the types of growth substrates. Two different Na-containing glass substrates, soda lime (SL) and alkali aluminosilicate (AA), were used. When the growth temperature was set to 600 °C, which exceeded the glass transition temperature $T_g$ of 570 °C for SL but lower than $T_g$ of 630 °C for AA, Na-rich Na–Mo–O alloys were formed on SL, whereas MoO$_x$ compounds without noticeable Na content were formed on AA[1]. Vapor–liquid–solid (VLS) growth occurred on SL at 600 °C as the temperature surpassed the eutectic temperature for Na–Mo–O alloys, $T_m$ of ~500 °C. However, vapor–solid–solid (VSS) growth occurred on AA, where MoO$_x$ compounds were in solid states with $T_m$ of ~ 800 °C.

We set the vapor environment to a strongly S-rich (Mo-poor) condition with a relatively large amount of injection of (C$_2$H$_5$)$_2$S precursors, compared to Mo(CO)$_6$ precursors. Growth Substrates were first loaded into the furnace and then evacuated to 50 mTorr. Then, a mixture of Ar (600 sccm) and H$_2$ (4 sccm) was introduced while ramping the temperature to 600 °C. Mo(CO)$_6$ diluted in Ar (0.3 sccm) and (C$_2$H$_5$)$_2$S diluted in Ar (1.5 sccm) were introduced up to a total pressure of ~10 Torr. The temperature was maintained at 600°C for 16 h to enable continuous film formation (Figure 1-3 in the main text). For sequential VLS and VSS growth on SL substrates (Figure 4), the growth temperature was maintained at 600 °C for 6 h and then cooled to 500°C for 2 h under constant flow of the gases Ar, $H_2$, $Mo(CO)_6$, and $(C_2H_5)_2S$.

The partial vapor pressures of $Mo(CO)_6$ and $(C_2H_5)_2S$ in the growth chamber are estimated from the ratio of flow rates for all gases, including Ar and $H_2$. Here, the canisters containing Mo and S precursors are pressured with Ar, up to 1000 Torr. The vapor pressure of Mo and S precursors are low at room temperature ($P_{Mo(CO)6, vapor} \sim 0.17$ Torr, $P_{(C2H5)2S, vapor} \sim 38$ Torr)[2], thus the injected flows of gases, containing Mo and S precursors are highly diluted. In growth environment with the total pressure of ~ 10 Torr, the partial pressures of $Mo(CO)_6$ is estimated to $8.42 \times 10^{-7}$ Torr, which is lower than $9.51 \times 10^{-4}$ Torr for $(C_2H_5)_2S$ by several orders of magnitude. It can produce the growth environment to be highly Mo-poor (S-rich), as depicted in Figure 1c of the main text.

**Scanning transmission electron microscope (STEM) characterizations**

Monolayer $MoS_2$ was suspended on a Quantifoil TEM Cu-gird. STEM data were obtained using JEM-ARM 200CF (JEOL, Japan) equipped with a fifth-order spherical aberration corrector (ASCOR, CEOS GmbH, Germany) at the Materials Imaging & Analysis Center of POSTECH in South Korea. The acceleration voltage was set to 200 kV, and a high-angle annular dark-field detector (inner angle: 54 mrad, outer angle: 216 mrad) with 40-μm condenser lens aperture was used to acquire the STEM images.

**Tilt-boundary model**

The $d_{defects}$ along GBs, expected for a given $\theta_t$, is estimated using a tilt-boundary model[3]. The atomic mismatch at GBs is accommodated by defects of unique atomic structures with $\vec{b}$. If the defects are arranged in a row to form symmetric GBs, they can be aligned toward (1, 0) or (1, -2) directions of merging grains, resulting in zig-zag (ZZ) GBs or armchair (AC) GBs, respectively. In ZZ-GBs, a pair of a Mo 5|7 and a S 5|7 defect form an array, and the density of the pair is $d_{(1,1)} = \frac{2}{|\vec{b}_{(1,1)}|} sin\frac{\theta_t}{2}$, where $\vec{b}_{(1,1)}$ is the Burgers vector for the pair, and $\theta_t$ is the tilt angle between grains. Thus, the total density of 5|7 defects is $d_{(1,0)} = \frac{2}{\sqrt{3}} \cdot \frac{2}{|\vec{b}_{(1,0)}|} \cdot sin\frac{\theta_t}{2}$, where $\vec{b}_{(1,0)}$ is the Burgers vector for a single 5|7 defect. On the other hand, for AC-GBs aligned to (1, -2), $d_{(1,0)} = \frac{2}{|\vec{b}_{(1,0)}|} sin\frac{\theta_t}{2}$. Our data (Figure S5) follow the relation between $d_{(1,0)}$ and $\theta_t$ for AC-GBs.

**Shear strain analyses**

PRESTem program was performed to extract the coordinates of Mo atomic positions for the strain analysis in atomic scale[4]. AC direction of each grain is a long diagonal vector of a rhombus with adjacent 4 Mo atoms, and the other diagonal vector of a rhombus is defined as ZZ direction (Figure 3c in the main text). The shear strain is deduced by distortion angle, $k$, formed between two Mo–Mo diagonals of the rhombus in a grain.

**Density functional theory (DFT) calculations**

DFT calculations were performed as implemented in the Vienna *ab initio* simulation package (VASP) code[5]. The projector-augmented wave (PAW) pseudopotentials[6,7] and a kinetic energy cutoff of 500 eV were applied. The Perdew, Burke, and Ernzerhof (PBE) functional within the generalized gradient approximation (GGA) was used for the exchange correlation functional[8]. The calculated hexagonal lattice constant of monolayer MoS$_2$ was 3.183 Å in PBE. In order to obtain the intrinsic properties of isolated GB defects, we constructed a GB loop model in a 6×6×1 monolayer MoS$_2$ supercell (Figure S10), through which we could avoid GB dipole effect (in a supercell with both polar GBs) and edge effect (in a finite size MoS$_2$ supercell with a GB and cluster edges). Vacuum layer thickness was 6.4 Å. The Γ k-point in the 6×6×1 supercell was used for the Brillouin zone summation. The local atomic configurations were relaxed to less than 0.001 eV/Å in the Hellmann–Feynman forces. The formation energies ($E_{form}$) of isolated GBs (Figure 1b in the main text) are calculated as $E_{form} = \{E_{tot} - (N_{Mo}\mu_{Mo} + N_S\mu_S)\}/N_{core}$, where $E_{tot}$ is the total energy of the supercell containing the GBs, $N_{Mo}$ and $N_S$ are the numbers of Mo and S atoms in the supercell, respectively, $\mu_{Mo}$ and $\mu_S$ are the chemical potentials of the Mo and S, respectively, and $N_{core}$ is the number of GB cores in the supercell. In our case, $N_{core}$ = 3. The $\mu_{Mo}$ and $\mu_S$ satisfy the relation of $\mu_{Mo} + 2\times\mu_S = E_{tot}$(MoS$_2$), where $E_{tot}$(MoS$_2$) is the total energy of the pristine monolayer MoS$_2$ per a formula unit. In the Mo-rich limit condition, $\mu_{Mo,max} = E_{tot}$(Mo), where $E_{tot}$(Mo) is the total energy of a Mo bcc metal per an atom, and then $\mu_{S,min} = (E_{tot}(MoS_2) - \mu_{Mo,max})/2$. In the S-rich limit condition, the S orthorhombic crystal is chosen as a S reservoir, and $\mu_{S,max} = E_{tot}(S)$, where $E_{tot}(S)$ is the total energy of the S orthorhombic crystal per an atom. In the thermodynamically allowed range of $\mu_S$, $\mu_{S,min} < \mu_S < \mu_{S,max}$, we plot the $E_{form}$ as a function of $\mu_S - \mu_{S,min}$ in Figure 1b. In calculation of DOS (Figure 4), we used 6×6×1 k-point mesh in the 6×6×1 supercell.

**Analyses of photoluminescence (PL) spectra in polycrystalline MoS$_2$ films**

To estimate the electron concentration $n$ from the measured PL spectra (Figure S9) in polycrystalline MoS$_2$ films, we use a three-level model for the relaxations of excited particles; it considers trions, excitons, and the ground state[9]. The ratio of spectral weights $w$ between the trion peak, $I_{tr}$, and the exciton peak, $I_{ex}$, is given as $I_{tr}/I_{ex}$. It is proportional to $\frac{\gamma_{tr}}{\gamma_{ex}} \times \frac{N_{tr}}{N_{ex}}$, where $\gamma_{tr}$ ($\gamma_{ex}$) presents the radiative decay rate of the trion (exciton), and $N_{tr}$ ($N_{ex}$) presents the concentration of the trion (exciton) at the equilibrium condition. $\frac{N_{tr}}{N_{ex}}$ is determined by $\frac{k_{tr}(n)}{\Gamma_{tr}}$. $k_{tr}(n)$ is the free electron concentration, $n$-dependent parameter that is equivalent to $\alpha n$, where $\alpha$ is the parameter for transition from excitons to trions[9,10]. $\Gamma_{tr}$ is the total decay rate of radiative and non-radiative components of the trion. Therefore, $w$ is simplified to $An$, where $A = \frac{\gamma_{tr}}{\gamma_{ex}} \times \frac{\alpha}{\Gamma_{tr}}$. We use the $A$ value from a reference[11]. As results, $n$ for VLS-grown films and VSS-grown films were deduced as 6.7 × 10$^{12}$ cm$^{-2}$ and 1.1 × 10$^{13}$ cm$^{-2}$, respectively.

The areal density for defects is roughly estimated by considering the total length of GBs in a unit area, and the $d_{defects}$ along the GB. Therefore, the areal density is equal to $(3L/2) \times d_{defects} \times (1/S)$, where $L$ is the average length of the edge of a grain, and $S$ is the average area of grain, which is equivalent to $\sqrt{3}/4 \times L^2$. The $L$ is measured as 0.26 μm for the VSS-grown films[1]. Assuming random $\theta_t$ within the range from 0° to 34°, we use the average $d_{defects}$ value of 0.85 nm$^{-1}$, which corresponds to $\theta_t \sim 17°$ (see Figure S5). We estimate the number of total defects to ~1.12 × 10$^{12}$ cm$^{-2}$ in the VSS growth. The ratio of S 5|7 defects is measured about 40 % for VSS-growths (see Figure 2e in the main manuscript). Therefore, the number of S 5|7 defects is ~4.5 × 10$^{11}$ cm$^{-2}$. While this value is smaller than the $n$ difference between VLS- and VSS-films of 4.3 × 10$^{12}$ cm$^{-2}$,

which was deduced from the PL measurements, this discrepancy could originate from the uncertain value of the parameter $A$, which is associated with the dynamics of excited states and can vary strongly based on the materials[11,12].

**Structural analyses of defects**

The average length of homoelemental bonding (HB) in 5|7 defects, $d$ are 2.35 Å for Mo 5|7 defects and 1.38 Å for S 5|7 defects (Figure S7). The $d$ for Mo–Mo HB is much larger than the $d$ for Mo–S bonding of ~1.83 Å in the hexagonal lattice of $MoS_2$ crystals (projected from the top view), and the $d$ for S–S HB is much shorter than the $d$ for Mo–S bonding. As result, the hexagonal lattices nearby HB are distorted, making the average outer angle of defects, $β$ to be 126.8° for Mo 5|7 defects and 106.9° for S 5|7 defects. In theory[13], $d$ of Mo-Mo in Mo 5|7 and S-S bonds in S 5|7 defects were calculated as 2.08 and 1.93 Å, respectively. The values for $β$ were calculated as 125° and 116° for Mo 5|7 and S 5|7 defects, respectively. The theoretical values are somewhat different from the experimentally measured values, but the trend is similar, as both $d$ and $β$ are higher in Mo 5|7 defects. The $β$ value close to 120° for the perfect hexagonal lattice and large $d$ value in Mo 5|7 defects indicate that the strength for Mo-Mo HB is weak. One reason for the weak bonding is the lower coordination number, $CN$ of 5 for Mo atoms in the Mo–Mo HB than that for Mo atoms in the Mo–S bonding of the perfect crystal ($CN = 6$). The weak Mo–Mo HB makes the GBs with Mo 5|7 defects highly mobile, prompting structural reconstructions, as shown in Figure 3 of the main manuscript.

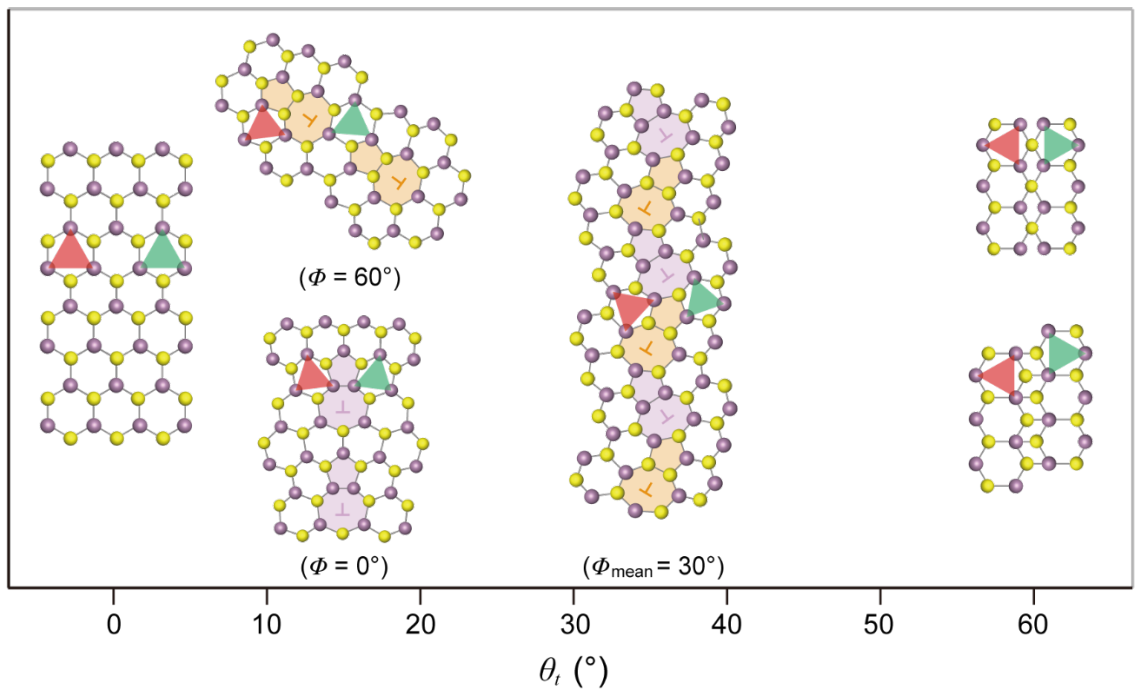

**Figure S1**. Schematics for atomic configuration of defects at tilt GBs as a function of $\theta_t$. The purple (yellow) spheres represent Mo (S) atoms.

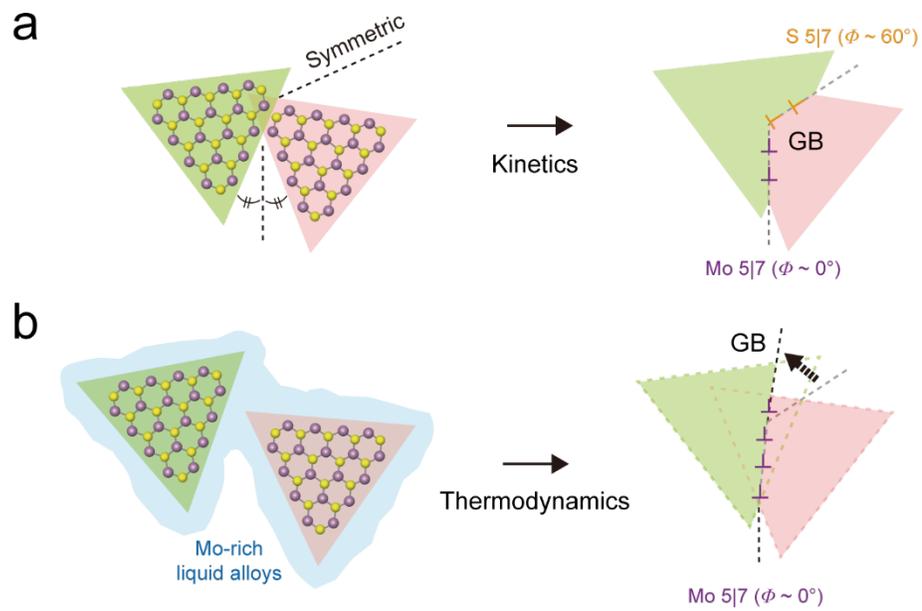

**Figure S2**. Schematics for continuous growth by 'zips-up' processes between adjacent tilted grains. (a) The GB structures are determined by kinetics in VSS growth and (b) they can be thermodynamically determined in VLS growth.

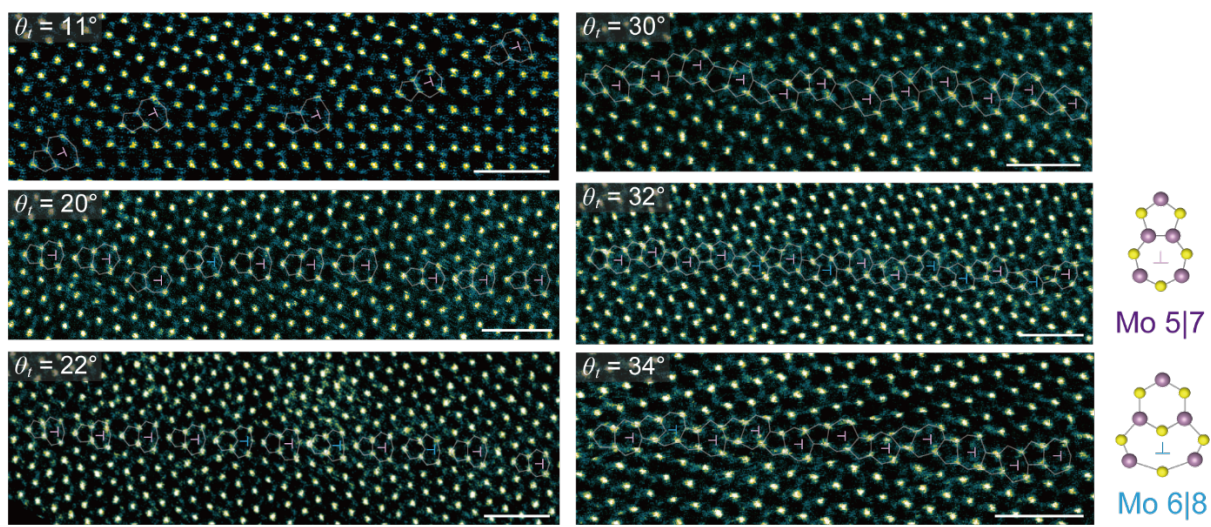

**Figure S3.** STEM atomic images over extended areas of GBs grown by VLS growth at various $\theta_t$. Regardless of $\theta_t$, most of the defects at GBs are Mo 5|7 defects with few Mo 6|8 derivatives. Scale bar: 1 nm.

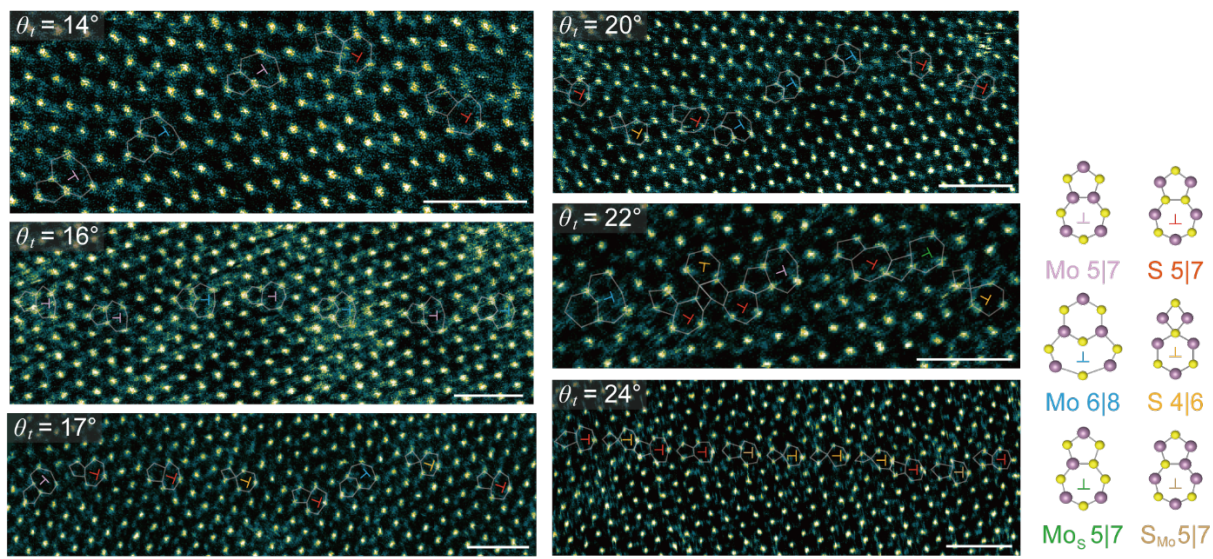

**Figure S4**. STEM atomic images over extended areas of GBs grown by VSS growth at various $\theta_t$. Various types of defects present at the GBs, including Mo defects with $\Phi \sim 0°$ (Mo 5|7, Mo 6|8, Mo$_S$ 5|7, Mo 4|6 defects) and S defects with $\Phi \sim 60°$ (S 5|7, S$_{Mo}$ 5|7, S 4|6). The various derivatives emerge by atomic reconstruction (Figure 2f, 2g in the manuscript) upon changes of $\mu_s$; this result suggests that the growth environments vary over time and space. Scale bar: 1 nm.

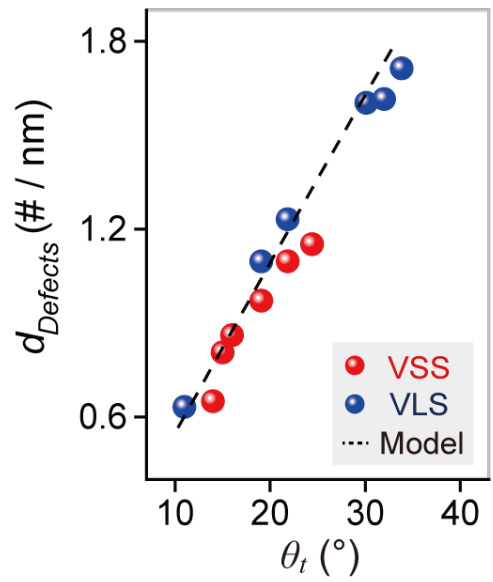

**Figure S5.** Experimentally measured $d_{defects}$ as a function of $\theta_t$ in VSS-grown (red sphere) and VLS-grown (blue sphere) films. The dotted lines indicate the theoretical model for symmetric AC-GBs.

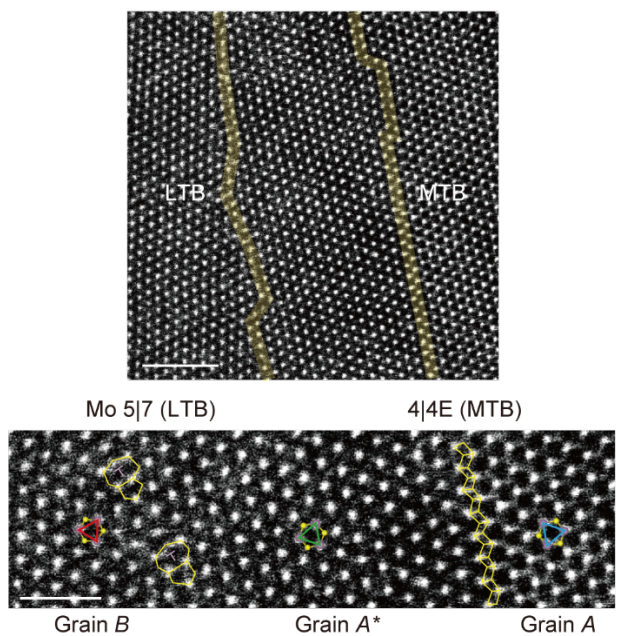

**Figure S6**. STEM atomic image of double GBs with LTB and MTB. It is another area from Figure 3b in the manuscript after deformation-twinning process. (upper image) Scale bar: 2 nm. (lower image) Scale bar: 1 nm.

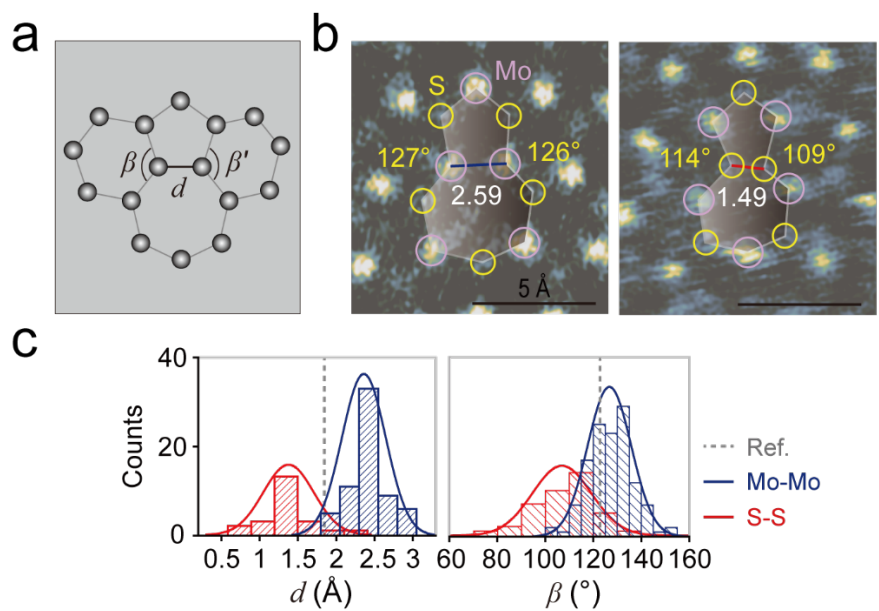

**Figure S7**. Structural analyses of Mo 5|7 and S 5|7 defects. (a) Atomic model of a 5|7 defect. Depending on the type of shared bond between 5 and 7 rings, the bond length, $d$ and bond angle, $\beta$ change. (b) Representative images of defects for 5|7 defects with Mo–Mo and S–S HB. $d$ and $\beta$ differ depending on the types of HB at the defects. (c) Histograms and best-fit normal curves of distributions of $d$ and $\beta$ for Mo–Mo and S–S HB; gray dotted lines: $d$ of Mo–S bonds (projected from the top view), and $\beta$ of 123.4° in the atomic model.

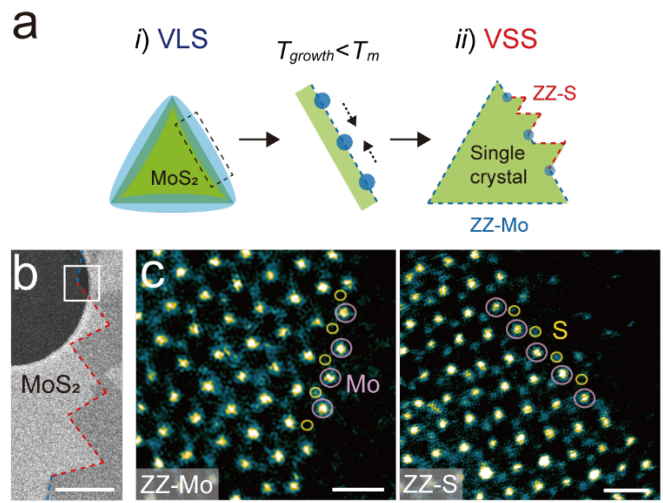

**Figure S8.** (a) Schematics for the two-step growth with sequential VLS and VSS growth modes controlled by the growth temperature. (b) TEM image of a single-crystal domain with sawtooth-like edges. Scale bar: 500 nm. (c) Zoomed-in STEM images of the white box in Figure S8b; left image: close up of blue dotted edge in S8b, which shows a ZZ–Mo edge; right image: close up of red dotted edge in Figure S8b, which shows a ZZ–S edge. Scale bars: 0.5 nm.

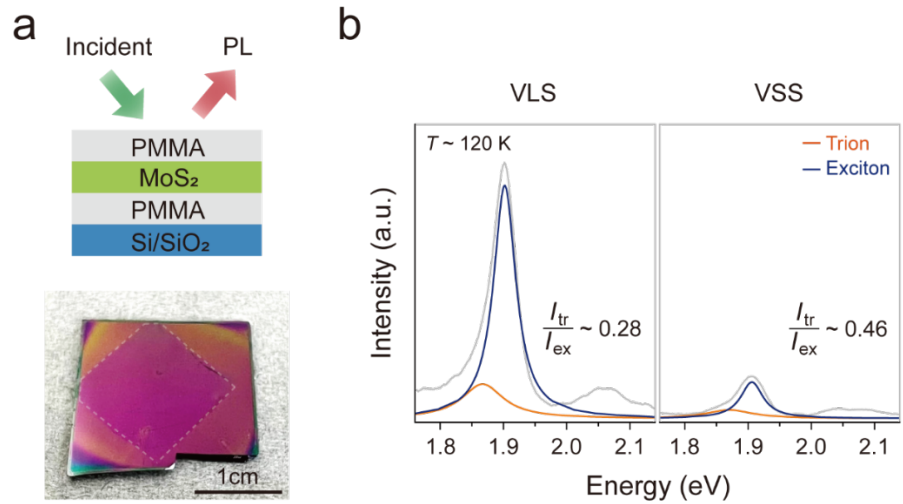

**Figure S9.** (a) Schematics of the structure of samples for PL measurements; inset image: photograph of a centimeter-scale $MoS_2$ film. The films were encapsulated with polymethylmethacrylate (PMMA) to minimize the effects of interfaces on the PL spectra because $MoS_2$/PMMA interfaces are usually trap-free[14]. (b) Representative PL spectra of continuous films by VLS and VSS growth; orange lines: trion peaks; blue lines: exciton peaks. The PL spectra were measured in a vacuum at 120 K at a low incident power of ~ $10^3$ W/cm$^2$ using a 532-nm laser.

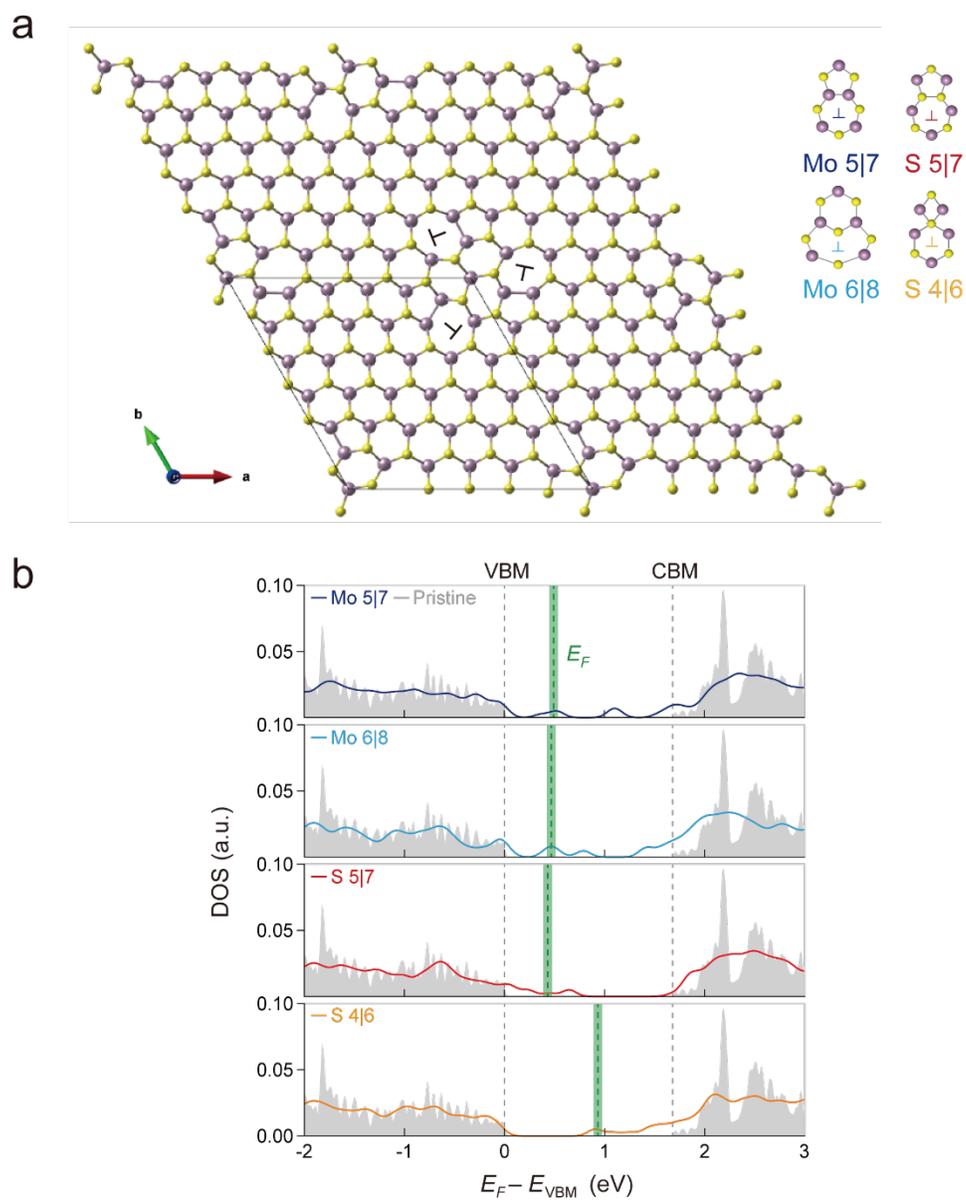

**Figure S10**. (a) GB loop model with defects, inclined to different directions for DFT calculations. (b) Electronic band structures of atomic defects embedded in monolayer $MoS_2$ crystals by DFT calculations.

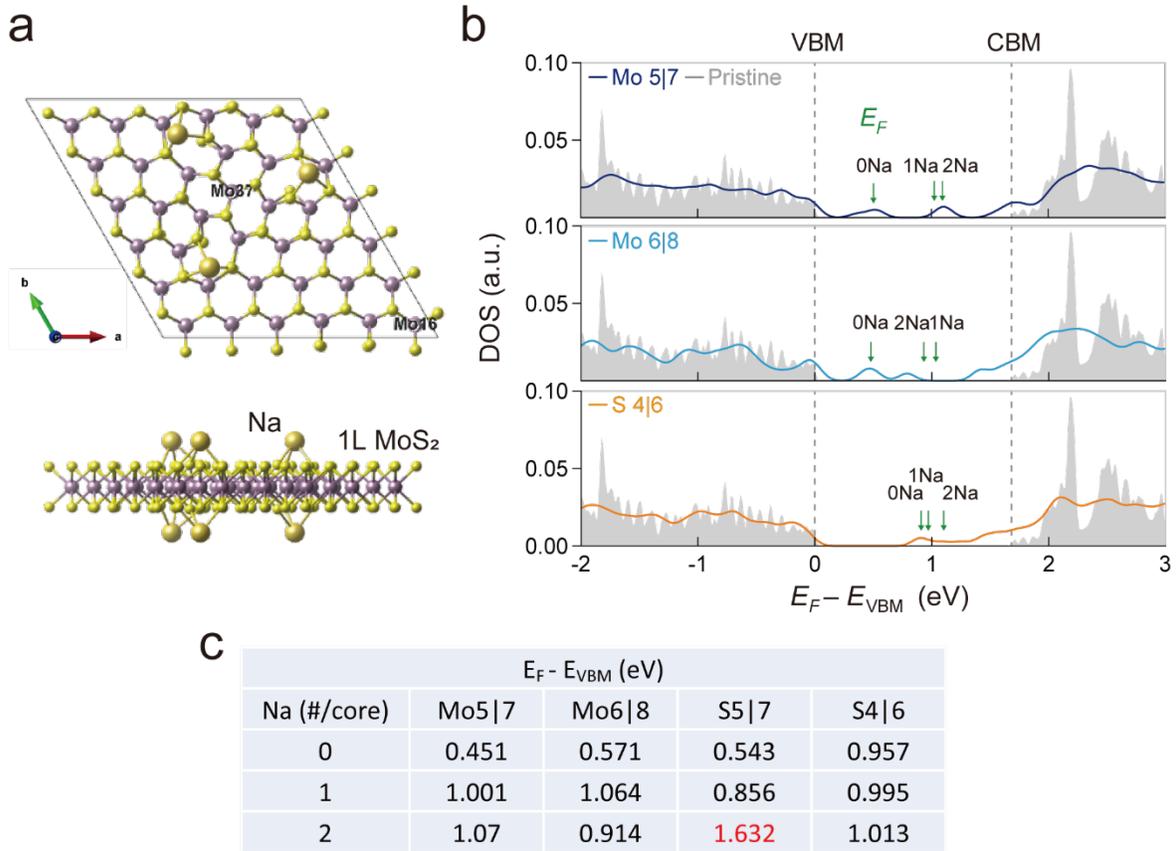

**Figure S11**. (a) Top and side view of unit cell used in the calculation, based on GB loop model with Na adsorption. (b) Deep acceptor levels of Mo 5|7, Mo 6|8, and S 4|6 defects, which should be occupied by electrons in n-type $MoS_2$, originated from Na donors, are indicated by the arrows. (c) Calculated Fermi levels of the Na-adsorbed defects with respect to valence band maxima of the pristine $MoS_2$ lattice, depending on the number of adsorbed Na atoms. The values of S 5|7 defect in the table are presented in Figure 4g of the main manuscript.